\documentclass[showpacs,showkeys,superscriptaddress,preprintnumbers,%
nofootinbib,amssymb,amsfonts,twoside,a4paper]%
{revtex4}

\usepackage[dvips]{graphicx}

\newcommand{\ie}{\emph{i.e.,} }
\newcommand{\cf}{\emph{cf.}~}
\newcommand{\fig}[1]{Fig.~\ref{#1}}

\def\Eq#1{Eq.~(\ref{#1})}
\def\Eqs#1{Eqs.~(\ref{#1})}
\def\eq#1{(\ref{#1})}
\def\sect#1{Sect.~\ref{#1}}

\newcommand{\rmd}{\mathrm{d}}

\def\on{\rho}
\def\step{n}
\def\ntot{{\cal N}}

\def\IT{non-equilibrium thermodynamics }
\def\VolContr{\varsigma}
\def\iirr{\Sigma^{\rm (irr)}}
\def\specent{P}

\def\rstar{\on^{\star}}
\def\Pe{{\rm Pe}}
\def\Esc{\kappa}
\def\Decay{\gamma}

\begin{document}
\date{December 14, 2001}
\title{Escape-rate formalism, decay to steady states, \\
  and divergences in the entropy-production rate}

\author{J\"{u}rgen Vollmer} 
\email{vollmer@mpip-mainz.mpg.de} %
\homepage{http://www.mpip-mainz.mpg.de/~vollmer} %
\affiliation{Max-Planck-Institute for Polymer Research,
  Ackermannweg 10, 55128 Mainz, Germany} 
\author{L\'aszl\'{o} M\'aty\'as} 
\affiliation{Institute for Theoretical Physics, 
  E\"{o}tv\"{o}s University, 
  P. O. Box 32, 
  H-1518 Budapest, Hungary}
\author{Tam\'as T\'{e}l} 
\email{tel@poe.elte.hu} %
\affiliation{Institute for Theoretical Physics, 
  E\"{o}tv\"{o}s University, 
  P. O. Box 32, 
  H-1518 Budapest, Hungary}

\begin{abstract}
\ \\
\rule{18mm}{0mm}
\parbox{10cm}{\rule{10cm}{0.1ex}\\[2mm]
  In summer 1997 we were sitting with Bob Dorfman and a few other
  friends interested in chaotic systems and transport theory on a
  terrace close to Oktogon in Budapest. While taking our (decaf)
  coffee after a very nice Italian meal, we discussed about
  logarithmic divergences in the entropy production of systems with
  absorbing boundary conditions and their consequences for the
  escape-rate formalism. It was guessed at that time that the problem
  could be resolved by a careful discussion of the \emph{physical}
  content of the absorbing boundary conditions. To our knowledge a
  thorough analysis of this long-standing question is still missing.
  We dedicate it hereby to Bob on occasion of his 65th birthday.\\
\rule{10cm}{0.1ex}}
\end{abstract}

\pacs{05.70.Ln, 05.45.+b, 05.20.-y, 51.10.+y} %
\keywords{\parbox[t]{10cm}{
    Entropy production,
    deterministic chaos, absorbing boundary condition,\\ 
    escape-rate formalism, multibaker maps}}

\preprint{submitted to \emph{J.\ Stat.\ Phys.}}

\maketitle
\pagestyle{myheadings}
\markboth{{\sc J. Vollmer, L.\ M\'aty\'as,} and {\sc T.\ T\'el}}%
{Escape-rate formalism, steady states, and entropy production}

\section{Introduction}
\label{sec:intro}

The \emph{escape-rate formalism}
\cite{GasNic90,DorfGas95,BTV96,GaspardBook99} aims at identifying
transport coefficients based on the asymptotic decay rate of an
initial non-stationary density profile towards an empty steady state
selected by absorbing boundary conditions where all particles
disappear from the interior of the system. In this formulation the
relaxation problem has widely been studied in the context of Markov
chains (\cf the sections on \emph{survival probabilities} in
\cite{HauKeh87}, and on \emph{absorbing states} in \cite{BreBook99}),
as well as for deterministic chaotic systems
\cite{Tel90,OttTel93,Tel96}.  However, the choice of an empty
asymptotic state places severe constrains on the formalism.  In
particular, the thermodynamic entropy production picks up
logarithmically diverging contributions at the boundaries.  After all,
a vanishing density is physically unrealistic.  In particular, it
leads to the breakdown of the concept of an entropy-production density
even in the framework of classical irreversible thermodynamics.
Close to a boundary at $x=0$ the absorbing boundary condition
requires a density profile  of the form $\on(x)=\alpha x$.
For purely diffusive particle transport, a neighborhood of size
$\Delta$ gives then rise to the entropy production%
\footnote{
The Boltzmann constant is taken to be unity throughout this paper.}
\begin{equation}
   \Sigma^{\rm (irr)} (\Delta)
   \equiv
   \int_{0}^{0+\Delta} \rmd x \; D
        \left( \frac{\partial_x \rho}{\rho} \right)^2 \rho
   =
   D \alpha \;
      \lim_{\delta\to 0} \ln\frac{\Delta}{\delta} .
\label{eq:1}
\end{equation}
In the present paper this logarithmic divergence will be discussed
from the point of view of spatially extended chaotic systems whose
transport properties fully agree with the predictions of irreversible
thermodynamics. To keep the calculations as transparent as possible,
the discussion is given for isothermal multibaker maps. However, the
described picture should apply in general.

In \sect{sec:EscRatRevisit} \Eq{eq:1} is contrasted with the
prediction of the irreversible entropy changes in an entropy based on
the conditional density characterizing the chaotic saddle forming the
backbone of transport in the system.  This prediction always yields
finite values.
The isothermal multibaker map is introduced in \sect{sec:MBaker},
where also its entropy balance is worked out.  \sect{sec:Modes} deals
with the normal modes of the coarse-grained time evolution.
This allows us to address in \sect{sec:Balance} the origin of the
logarithmic divergences in the entropy production from the point
of view of a microscopic reversible dynamics.
The analysis makes use of the eigenvalues of the time-evolution operator
\cite{GasPRE96,GaspardBook99}, which does not depend on the nature
of the asymptotic state. To underline this observation, we also
discuss asymptotic states with uniform nonzero densities,
and point out how the divergences are lifted by an
arbitrarily small background density.
The discussion also shows why the changes of the irreversible entropy
based on the conditional density differ from the thermodynamic
expectation.  The presence of an arbitrarily small background density
in the asymptotic case turns the entropy production to be finite, but
different from the prediction based on the conditional density.
In the concluding \sect{sec:Discussion} these findings are
complemented by a discussion of the behavior of the entropy
production in systems relaxing towards a typical nonempty
steady state of finite density as compared to the case of 
small (or even vanishing) background densities addressed in the main part.


\section{Entropy production based on conditional invariant measures revisited}
\label{sec:EscRatRevisit}

One of the early studies of thermodynamic entropy production in
deterministic dynamical systems was based on the escape-rate formalism
introduced by Gaspard and Nicolis \cite{GasNic90}.  As a
generalization of it, Breymann, T\'{e}l and Vollmer \cite{BTV96}
considered open dissipative dynamical systems in continuous time.  To
characterize their irreversible features, they suggested to use the
entropy
\begin{equation}
   s(t) = - \int \rmd x \; \psi^{(t)}(x) \ln \psi^{(t)}(x) 
\label{eq-s}
\end{equation}
based on the \emph{normalized conditional phase-space density}
$\psi^{(t)}(x)$, describing the probability to find a point which
has not yet escaped the system by time $t$ at phase-space
coordinate $x$.
Because $\psi^{(t)}(x)$ is a single-particle property,
$s(t)$ can be considered as a specific entropy
(total entropy per number of particles).
The initial condition is selected from an
arbitrary \emph{smooth} distribution $\psi^{(0)}(x)$.
As time goes on, the phase-space volume of $\psi^{(t)}$ is
exponentially shrinking as $\exp{[-\VolContr(x)\, t]}$, where the local
phase-space contraction rate $\VolContr(x) = \sum \lambda_i(x) - \Esc$ is
a smooth function of the coordinates \cite{Tel90}.  Here, $\Esc$ is the
escape rate from the system, and $\lambda_i(x)$ are the local Lyapunov
exponents that describe the phase-space contraction in the independent
directions $i$ in phase space (\cf\cite{Tel90,OttTel93} for detailed
discussions of conditional invariant densities and Lyapunov exponents).
On its support the value of the conditional density $\psi^{(t)}(x)$
is exponentially increasing due to its normalization.
More precisely, it increases like
   ${\psi^{(t+\rmd t)}(x)}
    = \exp{[\VolContr(x)\,\rmd t]}\;{\psi^{(t)}(x)}\;\chi^{(t+\rmd t)}(x)$,
where $\chi^{(t+\rmd t)}$ is the characteristic function of the support at
time $t+\rmd t$. The entropy $s(t+\rmd t)$ at time $t+\rmd t$ can be
determined by inserting this relation into \Eq{eq-s}:
\begin{eqnarray}
   s(t+ \rmd t)
& = &
   - \rmd t \int\rmd x \;
      \VolContr(x) {\psi}^{(t)}(x) \; e^{\VolContr(x)\,\rmd t} \chi^{(t+\rmd t)}(x)
\nonumber\\[2mm]
  &&
   - \int\rmd x \;
      {\psi}^{(t)}(x) \ln\left[{\psi}^{(t)}(x) \chi^{(t+\rmd t)}(x)\right] \;
         e^{\VolContr(x)\,\rmd t} \chi^{(t+\rmd t)}(x) .
\label{eq-ds}
\end{eqnarray}
In both integrals the decrease of the support of $\psi$ is
counterbalanced by the factor $\exp[\VolContr(x)\rmd t]$. The
first integral is the phase-space average $\bar{\VolContr}$ of
$\VolContr(x)$.  For $\rmd t \to 0$ the second one tends to the
specific entropy $s(t)$ at time $t$. Hence, in the long-time limit
the time derivative of the entropy
\begin{equation}
   \frac{d s}{d t} = - \bar{\VolContr} =  \sum_i \bar{\lambda_i}  - \Esc ,
\label{eq-sig}
\end{equation}
is the difference of the sum of the average Lyapunov exponents
$\bar{\lambda_i}$ on the saddle, and the escape rate $\Esc$ from
the saddle.
The average is taken with respect to the density $\psi(x)$ of the
conditionally-invariant measure.  This measure is time independent.
Its support is the unstable manifold of the chaotic saddle, \ie the
union of the never escaping orbits in the system.  The fact that the
time derivative approaches a constant reflects the ever refining
fractal structures in the density due to the chaoticity of the
dynamics.

We now compare $s(t)$ with a coarse-grained entropy
$s^{\rm(cg)}(t)$ computed in an analogous way from a coarse-grained
conditional density $\psi^{\rm(cg)}(t)$, which --- in contrast to
$\psi$ --- does converge towards a stationary distribution. The 
irreversible entropy production is then obtained as (\cf\cite{BTV96})
\begin{equation}
   \specent^{\rm(irr)}
\equiv
   \frac{d }{d t} \left[ s^{\rm(cg)}(t) - s(t) \right]
\stackrel{\mbox{\tiny long times}}{\longrightarrow}
   \Esc - \sum_i \bar{\lambda_i} .
\label{eq:sDotIrr}
\end{equation}
It measures the lack of information on the microscopic state due to
the finite resolution of the coarse-grained description. Similarly to
$s$, $\specent^{\rm(irr)}$ is a specific quantity.

In systems with a reversible dynamics the phase-space contraction is
proportional to the displacement in the direction of an applied field
\cite{EvaMorBook90,CELS:PRL,VTB98,BTV98,TelVol00,VollmerHabil}.
Consequently, in an open system with reversible dynamics the sum of
the average Lyapunov exponents on the chaotic saddle is zero since the
average number of steps in the direction of positive phase-space
contraction is the same as in the opposite.  Therefore, its specific
irreversible entropy production amounts to the escape rate,
\begin{equation}
\specent^{\rm (irr)} = \kappa .
\end{equation}

In the following we revisit this argument in the light of recent
developments \cite{VTB97,VTB98} dealing with steady states instead of
empty asymptotic states. We work out the irreversible entropy
production for an isothermal multibaker map with reversible
microscopic dynamics subjected to absorbing boundary conditions.

\section{The isothermal multibaker map}
\label{sec:MBaker}

\begin{figure}
\[ \includegraphics{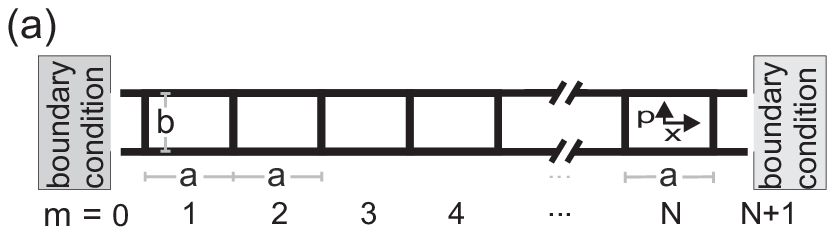} \]
\[ \includegraphics{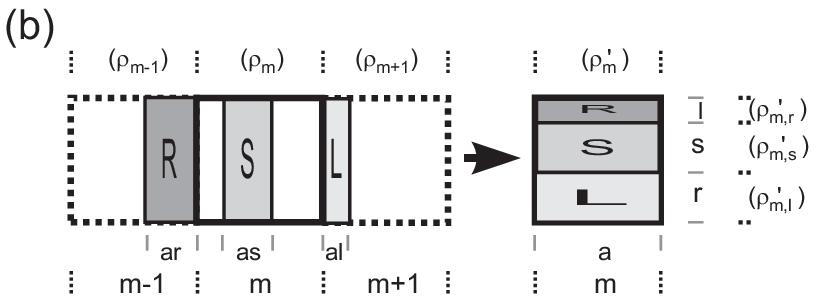} \]
\caption[]{
Graphical illustration of the action of the multibaker map of
length $L=aN$ on the phase space $(x,p)$ over a time unit $\tau$.
(a) The mapping is defined on a domain of $N$ identical
rectangular cells of size $a \times 1$, with boundary condition
imposed in two additional cells $0$ and $N+1$.
(b) The action of the map in any of the cells over time unit
$\tau$ is illustrated by the deformation of the labels $R$, $S$
and $L$ in the three branches of the map ending up in cell $m$.
The average value of the density on the cells (strips)
[\cf\Eq{eq:nm'}] is given on the margins.
\label{fig:MBaker}}
\end{figure}

Multibaker maps model particle transport in spatially extended systems
by a chain of mutually interrelated baker maps
\cite{GaspardJSP92,Tasaki95,TVB96,VTB97,VTB98,BTV98,Gent,GasPhysA97,GilDor99,TasGas99}.
They consist of $N$ identical cells of width $a$ and height $1$ (the
phase-space) in the $(x,p)$ plane.  The cells are labeled by the index
$m$ (Fig.~\ref{fig:MBaker}a). After each time unit $\tau$, every cell
is divided into three columns (Fig.~\ref{fig:MBaker}b).  Here we
consider the case when the right (left) column of width $a r$ ($a l$)
is mapped onto a strip of width $a$ and of height $l$ ($r$) in the
right (left) neighboring cell. The middle one, which is of width $a
s$, preserves its area, such that its image attains a height $s$, and
$r+l+s=1$.  There are more general parameter settings conceivable, but
earlier work \cite{VTB97,VTB98,VollmerHabil} showed that the
associated macroscopic behavior is then not compatible with
irreversible thermodynamics.

The dynamics of the multibaker map models a microscopic dynamics
described in the single-particle phase space.  It is deterministic,
invertible, chaotic, and mixing \cite{Hopf37,TelVol00}. To describe
irreversible processes one follows the coarse-grained densities
$\on_m$ obtained by averaging over the cells \cite{VTB97,VTB98,BTV98}.
To emphasize the particular choice of coarse-graining over the cells,
the coarse-grained densities are also called the \emph{cell
  densities}.
The dynamics of the multibaker map is the same for all cells.  There
might  be inhomogeneities in the densities, but the evolution equations
are translation invariant.

\subsection{Evolution of the cell density}

In order to find results consistent with \IT we always consider
initial conditions with a uniform density 
in every cell $m$. This is convenient from a technical point of view,
and does not lead to a principal restriction of the domain of validity
of the model as discussed in \cite{VTB98,MTV:PRE00b,VollmerHabil}.
Under such conditions the parameters $r$ and $l$ can be considered as
transition probabilities from a cell to its right and left neighbor,
respectively. After one step of iteration the densities $\on_{m,i}'$
on the strips $i=R,S,L$ of cell $m$ are (\cf\fig{fig:MBaker}b)
\begin{eqnarray}
   \on_{m,r}'  =  \frac{r}{l} \; \on_{m-1} , \qquad
   \on_{m,s}'  =  \on_{m} , \qquad
   \on_{m,l}'  =  \frac{l}{r} \; \on_{m+1} .
\label{eq:nm'}
\end{eqnarray}
The factors
   ${r}/{l}$
and
   ${l}/{r}$
give rise to local contraction or expansion of the phase-space
volume. One of the factors is larger than unity, and
characterizes a local contraction, while the other gives rise to
an expansion.

Since in every cell the density remains uniform in the horizontal
direction, this update holds at all times such that the coarse-grained
density distributions $\on_m$ and
$ \on'_m=r\on_{m-1}+s\on_m+l\on_{m+1} $ %
at the respective times $n$ and $n+1$ are related by the master
equation
\begin{equation}
   {\on'_m}
=
   (1-r-l) \on_m + r \on_{m-1} + l \on_{m+1} .
\label{eq:rhoprime}
\end{equation}
Multiplying the equation by $\tau^{-1}$ and introducing the
current
\begin{eqnarray}
   j_{m}
=
   \frac{a}{\tau}(r\on_{m}-l\on_{m+1})
\label{eq:jm+}
\end{eqnarray}
through the right boundary of cell $m$, \Eq{eq:rhoprime} appears
in the form of the continuity equation
\begin{equation}
   \frac{\on'_m-\on_m}{\tau}
=
   -\frac{j_m-j_{m-1}}{a} .
\label{eq:dt_varrho'}
\end{equation}
The current through the left boundary of cell $m$ is the
same as the current flowing through the right boundary of cell $m-1$.

\subsection{Diffusion, drift, and the macroscopic limit}

The transition probabilities
   $r$ and $l$
govern the evolution of the coarse-grained density $\on_m$. In view of
the master equation \eq{eq:rhoprime}, the cell-to-cell dynamics of the
model is equivalent to the dynamics of an ensemble of random walkers with
fixed step length $a$ and local transition probabilities $r$ and $l$ over
time unit $\tau$.
In terms of the local drift $v$ and diffusion coefficient $D$
\cite{ReifBook} the transition probabilities $r$ and $l$ can be
expressed as
\begin{eqnarray}
   r = \frac{\tau D}{a^2} \left(1+\frac{a v}{2D} \right) , \qquad
   l = \frac{\tau D}{a^2} \left( 1-\frac{a v}{2D} \right) ,
\label{eq:rmlm}
\end{eqnarray}
such that the current appears in a form very close to its
thermodynamic counterpart, viz.
\begin{equation}
   j_m = \frac{v}{2} (\on_m + \on_{m+1})
      - D \; \frac{\on_{m+1} - \on_m}{a} .
\label{eq:jm+b}
\end{equation}

The \emph{macroscopic limit} expresses a separation of scales where
density gradients inside cells may be neglected, {while} density
differences between the cells and the temporal evolution {of the
  cell densities} are only taken into account in leading order. In
other words, in this limit $a \ll L=a N$, $\tau \ll L^2/D$, $L/v$, and upon
introducing the quasi-continuous spatial and temporal variables $x
\equiv a m$ and $t \equiv \tau \step$, the current $j_m$
[\Eq{eq:jm+b}] takes the macroscopic limit $j(x;t) = v \, \on(x;t) - D
\, \partial_x \on(x,t)$ , while \Eq{eq:dt_varrho'} becomes $ \partial_t \on(x;t) =
- \partial_x j(x;t) = - v \, \partial_x\on(x;t) + D \, \partial_x^2\on(x;t) $ .  The
macroscopic density evolves according to the advection-diffusion
equation.

\subsection{Local entropy balance}
\label{ssec:Balance}

The coarse-grained entropy of cell $m$ is defined as
\begin{equation}
   S_m = - a  \on_m \ln\frac{\on_m}{\rstar} .
\label{eq:Sm}
\end{equation}
It fulfills a local entropy balance in direct analogy to the one of
irreversible thermodynamics \cite{VTB98,BTV98}. In this equation
$\rstar$ is a constant reference density that is introduced for
dimensional reasons. In classical physics it expresses the free
choice of the origin of the entropy scale.

To derive the balance equation for \eq{eq:Sm} one identifies at any
given time the difference $S_m-S^{(G)}_m$ of the coarse-grained and
the Gibbs entropy $S^{(G)}_m$ as the information on the microscopic
state of the system which cannot be resolved in the coarse-grained
description. The Gibbs entropy is an analogous expression to
(\ref{eq:Sm}) given in terms of the non-coarse-grained phase-space
density $\on$, as $- \on \ln{\on}/\on^*$ integrated over the cell. For
a coarse-grained initial distributions the entropies coincide
initially (\ie $S_m = S^{(G)}_m$). After one time step the entropies
become (\cf Fig.~\ref{fig:MBaker}b)
\begin{equation}
   S_{m}^{(G)'}
=
- a  \; \left[
      s \on_m \ln\frac{\on_m}{\rstar}
    + r \on_{m-1} \ln\left( \frac{r}{l}\frac{\on_{m-1}}{\rstar} \right)
    + l \on_{m+1} \ln\left( \frac{l}{r}\frac{\on_{m+1}}{\rstar} \right)
\right]
\label{eq:SGprime}
\end{equation}
and
\begin{equation}
   S'_m = - a  \; \on'_m \ln\frac{\on'_m}{\rstar} .
\label{eq:Smprime}
\end{equation}

The temporal change of the lack of information is identified with
the irreversible entropy production $\Delta_i S_m$, and the change
$(S^{(G)'}_m-S^{(G)}_m)$ of the Gibbs entropy with the entropy
flux $\Delta_e S_m $. Thus, one obtains the discrete entropy
balance in any cell
\begin{equation}
   \frac{S'_m-S_m}{\tau}
 = \frac{\Delta_e S_m}{\tau} +\frac{\Delta_i S_m}{\tau}  .
\end{equation}

The form of the entropy production is [\cf~(\ref{eq:SGprime}) and
(\ref{eq:Smprime})]
\begin{eqnarray}
&&   \frac{\Delta_i S_m}{\tau}
 = 
  \frac{[S'_m-S^{(G)'}_m] - [S_m-S^{(G)}_m]}{\tau}
\nonumber\\
&& = 
 \frac{a}{\tau} \; \left[
   - \on_m' \ln\frac{\on_m'}{\on_m}
   + r \on_{m-1} \ln\left( \frac{r}{l} \frac{\on_{m-1}}{\on_m} \right)
   + l \on_{m+1} \ln\left( \frac{l}{r} \frac{\on_{m+1}}{\on_m} \right)
\right] ,
\label{eq:DiSm}
\end{eqnarray}
and the entropy flux becomes
\begin{eqnarray}
&&   \frac{\Delta_e S_m}{\tau} 
 = 
  \frac{S^{(G)'}_m - S^{(G)}_m}{\tau}
\nonumber\\
&& =   - \frac{a}{\tau}
\left[
   (\on_m^{'} - \on_m) \; \ln\frac{\on_m}{\rstar}
+
   r \on_{m-1} \ln\left( \frac{r}{l} \frac{\on_{m-1}}{\on_m} \right)
+
   l \on_{m+1} \ln\left( \frac{l}{r} \frac{\on_{m+1}}{\on_m} \right)
\right] .
\label{eq:DeSm}
\end{eqnarray}
In the macroscopic limit all expressions reduce to the respective
predictions of non-equilibrium thermodynamics
\cite{VTB98,BTV98,VollmerHabil}. Note that the entropy production
does not depend on the choice of the reference density $\rstar$.
The density of irreversible entropy production is then
${\Delta_i S_m}/{(a\tau)}$.

\subsection{Global entropy production}

The escape-rate formalism addresses the balance of the global entropy
of the chain.  The global coarse-grained entropy is
\begin{equation} \label{eq:totalentr}
S_{\rm tot} = \sum_{m=1}^{N} S_m = -a \sum_{m=1}^N \on_m
\ln \frac{\on_m}{\on^{\star}} .
\end{equation}
The associated global entropy production rate along
the chain is $\iirr = \sum_{m=1}^N \Delta_i S_m/\tau $,
and the total specific irreversible entropy production is obtained as
\begin{equation}
\specent^{\rm (irr)} =
\frac{\iirr}{{\cal N}} =
\frac{\sum_{m=1}^N \Delta_i S_m}{\tau {\cal N}}
\end{equation}
where ${\cal N}=\sum_{m=1}^N a \on_m$ is the number of particles in
the chain at time $n\tau$.  The total entropy production can be
rearranged to take the form
\begin{eqnarray}
\iirr &=& -\frac{a}{\tau}
         \sum_{m=1}^N \rho_m^{'} \ln \frac{\rho_m^{'}}{\rho_m}
         + \frac{a}{\tau}
         \left( r \rho_0 \ln \frac{r\rho_0}{l\rho_1}
              - l \rho_{N+1} \ln \frac{r\rho_N}{l\rho_{N+1}}
         \right)
\nonumber \\
&&        + \frac{a}{\tau} \sum_{m=1}^{N-1} (r\rho_m - l\rho_{m+1})
          \ln \frac{r\rho_m}{l\rho_{m+1}} ,
\label{eq:iirr1}
\end{eqnarray}
where $\rho_0$ and $\rho_{N+1}$ are the densities in the boundary cells.
It can conveniently be split into four terms
$\iirr = \iirr_t + \iirr_b + \iirr_d + \iirr_{\rm mix}$.
The first one
\begin{equation}
\iirr_t = - \frac{a}{\tau}
            \sum_{m=1}^{N}  \rho_m^{'} \ln \frac{\rho_m^{'}}{\rho_m}
\label{eq:iirrtt}
\end{equation}
is the contribution from the \emph{temporal} evolution of the density.
The contribution proportional to $\ln (r/l)$ of the last term in
(\ref{eq:iirr1})
contains the irreversible entropy production
\begin{equation}
\iirr_d = {\cal N} \frac{r-l}{\tau} \ln \frac{r}{l}
         \rightarrow {\cal N} \frac{v^2}{D}
\end{equation}
due to the presence of the \emph{drift} $v$.
It does not depend on the particular density distribution so that we
could immediately specify its macroscopic limit
(indicated by $\rightarrow$).
By means of (\ref{eq:jm+}) and (\ref{eq:dt_varrho'}) the remaining part
can be written as a sum of two terms. One of them,
\begin{equation}
\iirr_{\rm mix} = -\frac{a}{\tau} \sum_{m=1}^N (\rho_m^{'}-\rho_m)
              \ln\frac{\rho_m}{\rho^{\star}}
\label{eq:iirrmix}
\end{equation}
characterizes the contribution of \emph{mixing}
of the neighboring
densities. In order to arrive at this form the ratio
$\rho_m/\rho_{m+1}$ of the densities appearing at the right hand
side of (\ref{eq:iirr1}) was written as
$[(\rho_m/\rho^{\star})/(\rho_{m+1}/\rho^{\star})]$. The rest
\begin{equation}
\iirr_b = \frac{a}{\tau}
          \left[
            r \rho_0 \ln \frac{r\rho_0}{l\rho^{\star}} 
          + l \rho_{N+1} \ln \frac{l\rho_{N+1}}{r\rho^{\star}} 
          - l \rho_1  \ln  \frac{l\rho_1}{r\rho^{\star}} 
          - r \rho_N \ln \frac{r\rho_N}{l\rho^{\star}} 
          \right]
\label{eq:iirrbb}
\end{equation}
yields the \emph{boundary} contribution.
We shall be interested in the difference
between $\iirr$ and $\iirr_d$, called the irreversible
entropy production $\iirr_{\rm relax}$ connected to the \emph{relaxation}
process,
\begin{equation}
\iirr_{\rm relax} \equiv \iirr_t + \iirr_{\rm mix} + \iirr_b .
\label{eq:iirr2}
\end{equation}

\section{Normal modes of the coarse-grained time evolution}
\label{sec:Modes}

\subsection{Decaying modes and the steady state}

We are interested in the evolution of the density distributions
$\on^{(\step)}_m$ subjected to a fixed constant boundary condition
\begin{equation}
    \on^{(\step)}_{0} = \on^{(\step)}_{N+1} = \on_B
\label{eq:bc}
\end{equation}
at any time step $n$. Asymptotically, $\on^{(\step)}_m$ always
approaches the uniform density $\rho_B$. The time evolution of the
density can be explored by expanding the deviation
$\on_m^{(0)}-\on_B $ of the initial distribution from the
asymptotic state in terms of normal modes $\delta_m^{[\nu] (n)}$.
They vanish at both boundaries,
and only change in amplitude but not in their shape,
\begin{equation}
   \frac{\delta_{m}^{[\nu](\step+1)}}{\delta_m^{[\nu](\step)}}
=
   \exp (- \Decay_\nu \tau ) .
\label{eq:rhoRatio1}
\end{equation}
The integer $\nu$ labels different modes.
There are as many independent modes
as the number $N$ of the cells $\nu=1,\ldots,N$.
The normal modes take the respective forms
\begin{equation}
 \delta^{[\nu] (\step)}_{m}
\sim
   \exp\left( -\Decay_\nu \step \tau \right) \;
   \left( \frac{r}{l} \right)^{m/2} \;
   \sin\left( \frac{\pi \nu}{N+1} \, m \right) .
\label{eq:prob_mn}
\end{equation}
Substituting the ansatz into \Eq{eq:rhoprime} and rearranging the
trigonometric terms, one finds the decay rates
\begin{equation}
   \Decay_\nu
=
   - \frac{1}{\tau}
     \ln\left[ 1 - (r+l) + 2 \sqrt{rl} \;
     \cos\left( \frac{\pi \nu}{N+1} \right) \right]
\rightarrow
 \frac{\pi^2 \, D}{L^2} \, \nu^2  + \frac{v^2}{4D} .
\label{eq:gamma-k}
\label{eq:gamma-kN}
\end{equation}
For a general initial condition the \emph{asymptotic} decay is
  governed by the \emph{slowest} non-vanishing decay rate, $\Decay_1$.
  It coincides with the \emph{escape rate} $\Esc$ of the transiently
  chaotic motion \cite{OttTel93} inside the chain (i.e.,
  $\kappa\equiv\Decay_1$) \cite{GasNic90,GasDorf95,TVB96}.

The macroscopic limit of the decay rates has a
clear physical content. For vanishing $v$ it states that
relaxation is related to the typical diffusive decay rate $D/L^2$
of structures of size $L$. The factor $\pi^2$ characterizes the
geometry of the considered region (a band of width $L$ with
straight, parallel walls in our example). More complicated
geometries have been studied recently by Gaspard
\cite{GaspardJSP92}, and Kaufmann and collaborators
\cite{Kaufmann97,Kaufmann99}.

For a biased motion $v\neq 0$ the drift singles out one side of
the system and sweeps out the particles in that direction. This
mechanism dominates when the time $L/v$ to cross the system by the
biased motion becomes shorter than the typical time scale $L^2/D$
of diffusion, \ie for
\begin{equation}
    \Pe \equiv \left| \frac{v L}{D} \right|
\label{eq:CrossOver}
\label{eq:Pe}
\end{equation}
much larger than unity.
In the context of hydrodynamics, $\Pe$ is called  P\`{e}clet number
\cite{Liggett}. It measures the importance of diffusion relative to
advection. Strong diffusive effects are indicated by small P\`{e}clet
numbers. For fixed finite $v$ and $D$, the P\`{e}clet number is always large
for a sufficiently large system size $L$.

\subsection{Long-time relaxation and the slowest mode}

For sufficiently long times $n \gg 1$, the coarse-grained density
closely approaches the first normal mode.
Therefore, the density can be expressed as
\begin{equation}
   \on_{m}^{(\step)}
\equiv
   \on_B +
   \left(
      {\cal N}^{(\step)} - {\cal N}^{(\infty)}
   \right)\; \psi_m ,
\label{eq:rho_mn}
\end{equation}
where ${\cal N}^{(\infty)}=\rho_B L$ is the particle number in the
background which is also the asymptotic particle number in the system,
and
\begin{equation}
   \psi_m
=
 \frac{\cal A}{L} \left(\frac{r}{l} \right)^{m/2-(N+1)/4}
 \;\sin\frac{m \pi}{N+1}
\label{eq:psi-m-cg}
\end{equation}
is  the  \emph{coarse-grained} conditionally-invariant density.
It is normalized to unity
   ($1 \equiv a \; \sum_{m=1}^{N} \psi_{m}$),
by virtue of the normalization constant ${\cal A}$,
which is invariant under the exchange of $r$ and $l$.
Carrying out the summation of the complex geometric series defined
by (\ref{eq:psi-m-cg}) one finds
\begin{eqnarray}
{\cal A}  & = &
L\; \frac{1-\exp(-\kappa\tau)}{a \sqrt{rl}} \;
\frac{1}{\left(\frac{r}{l}\right)^{(N+1)/4}+\left(\frac{r}{l}\right)^{-(N+1)/4}} \;
  \frac{1}{\sin\frac{\pi}{N+1}}
\nonumber \\
& \rightarrow &
   \frac{\pi}{2} \;
   \frac{\left(\frac{\Pe}{2\pi}\right)^2+1}{\cosh\frac{\Pe}{4}}
= \kappa \frac{L^2}{2\pi D\cosh\frac{\Pe}{4}} .
\label{eq:ANPe}
\end{eqnarray}
Here, the relation
$ (r/l)^{N/4}=(1+av/D)^{N/4} \rightarrow \exp(\Pe/4) $ %
has been used to evaluate the macroscopic limit. The asymptotically
decaying density takes then the form
\begin{equation}
\on (x,t) =
\rho_B + \frac{\cal A}{L} \;
({\cal N}^{(t)} - {\cal N}^{(\infty)})\;
   \exp\left(\Pe\frac{2x-L}{4L}\right) \sin \frac{\pi x}{L} .
\end{equation}

\section{Boundary contributions to the irreversible entropy production}
\label{sec:Balance}

\subsection{Absorbing boundaries}
\label{ssec:SBalanceAbs}

In the case of a long-term relaxation towards an empty state
($\rho_0=\rho_{N+1}=\rho_B=0$), \Eq{eq:rhoRatio1} holds for the
full density $\rho_m$, and one can write
[see \Eq{eq:iirrtt}]
\begin{equation}
   \iirr_t
   \equiv
     \kappa {\cal N}  \exp( - \kappa \tau) = \kappa {\cal N}'
   \to \kappa \ntot,
\end{equation}
where $\ntot'$ is the number of particles at time $(n+1)\tau$.

The mixing term (\ref{eq:iirrmix}) can be expressed by means of the total
entropy (\ref{eq:totalentr}) to obtain
\begin{equation}
\iirr_{\rm mix}
   = \frac{1}{\tau} (e^{-\kappa\tau}-1) S_{\rm tot}
   \to - \Esc S_{\rm tot},
\end{equation}
and in view of $\rho_0=\rho_{N+1}=\rho_B=0$ the boundary
contribution (\ref{eq:iirrbb}) becomes
\begin{eqnarray}
\iirr_b
= & 
   - \frac{a}{\tau} \frac{\cal A \, N}{L} \sin \frac{\pi}{N+1}
& \textstyle \Biggl\{
   r \left( \frac{r}{l} \right)^{(N-1)/4}
 \ln \left[
  \frac{\cal A \, N}{L \rho^*} \left( \frac{r}{l} \right)^{(N+3)/4} \sin  \frac{\pi}{N+1}
     \right]
\nonumber\\ && \textstyle +
     l \left( \frac{r}{l} \right)^{-(N-1)/4}
 \ln \left[
  \frac{\cal A \, N}{L \rho^*} \left( \frac{r}{l} \right)^{-(N+3)/4} \sin  \frac{\pi}{N+1}
     \right]
\Biggr\}
\nonumber\\
= &  
   - \frac{a}{\tau} \frac{\cal A \, N}{L} \sin \frac{\pi}{N+1}
& \textstyle \Biggl\{
   \left[ r \left( \frac{r}{l} \right)^{(N-1)/4}
        + l \left( \frac{r}{l} \right)^{-(N-1)/4}
   \right] \;
 \ln \left[
  \frac{\cal A \, N}{L \rho^*} \sin  \frac{\pi}{N+1}
     \right]
\nonumber\\ && \textstyle +
   \left[ r \left(  \frac{r}{l} \right)^{(N-1)/4}
        - l \left(  \frac{r}{l} \right)^{-(N-1)/4}
   \right] \;
   \frac{N+3}{4} \log\frac{r}{l}
\Biggr\} .
\end{eqnarray}
Observing that in the macroscopic limit both $r$ and $l$ are in
leading order equal to $\tau D/a^2$, and that $\log(r/l)\to av/D$, one
obtains for the relaxation contribution to the total irreversible
entropy production
\begin{equation}
\iirr_{\rm relax} = \kappa {\cal N}
 \left[ 1 -  \frac{\Pe}{4} \tanh \frac{\Pe}{4} -
        \ln \left( \frac{{\cal N}}{L \rho^{\star}} \frac{\pi^2}{2}
                   \frac{1+(\Pe/2\pi)^2}{\cosh (\Pe/4)} \frac{a}{L}
            \right)
  \right]
        - \kappa S_{\rm tot} .
\end{equation}
This expression shows the expected logarithmic divergence since
$(a/L)\rightarrow 0$ in the macroscopic limit. On the other hand,
the result can  properly be interpreted only after evaluating
$S_{\rm tot}$. In particular, the reference density $\rstar$ has
to drop out again in the final result.

In the macroscopic limit the sum over $m$ in the definition of
$S_{\rm tot}$ becomes an integral. By using \Eqs{eq:rho_mn} one
finds
\begin{eqnarray}
  S_{\rm tot} 
&=& 
  - {\cal A \, N} \int_0^1 \rmd x \;
   \exp\left(\textstyle -\Pe\, \frac{2 x - 1}{4} 
   \right) \;
   \sin(\pi x)
\nonumber \\ && \qquad\qquad
   \ln \left[\frac{{\cal A \, N}}{L \rho^{\star}} \;
           \exp\left(-\Pe\,  \frac{2 x - 1}{4} 
           \right) \;
           \sin(\pi x) \right] .
\end{eqnarray}
Applying the relation (\ref{eq:ANPe}), we see that the specific entropy
\begin{equation}
 S_{\rm tot}
=
 {\cal N}  f \left(\Pe ,\frac{{\cal N}}{L\rho^{\star}}\right)
\end{equation}
is a function of the P\`{e}clet number and of the ratio of the
average density in the system ${\cal N}/L$ and the reference density
$\rstar$, \ie of another dimensionless constant 
that involves the parameter $\rstar$ selecting the origin of the
entropy scale (which, as mentioned earlier, is an arbitrary number in
classical physics). The total entropy cannot be evaluated exactly.
However, to obtain its behavior in leading order for very large and
small P\`{e}clet numbers it is sufficient to approximate the
expression under the logarithm by its maximum value.  In the two
limiting cases one thus finds in leading order in ${\cal N}$
\begin{equation}
    S_{{\rm tot}} =
\left\{
\begin{array}{lrl}
 -  {\cal N} \ln \left( \Pe \frac{\cal N}{\rho^{\star} L } \right)
    &\qquad \mbox{for } & \Pe \gg 1 ,
\\[1mm]
  - {\cal N} \ln\left( \frac{{\cal N}}{\rho^{\star} L} \right)
    &\qquad \mbox{for } & \Pe\ll 1 .
\end{array}
\right.
\label{eq:Stot}
\end{equation}
This implies for the specific entropy production
\begin{equation}
  \specent^{\rm (irr)}
=
\frac{\iirr_{\rm relax}}{{\cal N}}
=
\left\{
\begin{array}{lrl}
  \kappa  \ln \left( \Pe^{-1}\, \frac{L}{a} \right)
 & \qquad {\rm for } & \Pe\gg 1 ,
\\[1mm]
  \Esc  \ln \frac{L}{a}
  & {\rm for } &  \Pe\ll 1 .
\end{array}
\right.
\end{equation}
The result shows the expected logarithmic divergence due to the
boundary terms. It should be considered as an example clearly showing
the inadequateness of global quantities for characterizing
thermodynamic properties [\cf~\Eq{eq:1}]. The reason for the breakdown
of the prediction \eq{eq:sDotIrr} lies in the fact that the argument
leading to this result focuses on the shrinking of the support of the
measure by assuming the smoothness of the distribution along the
unstable manifold. It thus entirely disregards that the density is
very inhomogeneously distributed as a consequence of the absorbing
boundaries.  It should also be noted that the result obtained for the
diffusive case $\Pe \ll 1$ is the analog of the thermodynamic
expression (\ref{eq:1}) since in this case $\rho = N\pi/(2L)
\sin(x\pi/L)$, such that the parameter $\alpha$ of \Eq{eq:1} takes the
value $\alpha={\cal N} \pi/(2L^2)$ and $D\alpha={\cal N}\kappa/2$.
The presence of factor $1/2$ is due to the fact that Eq.~(\ref{eq:1})
gives the contribution of one end only.

\subsection{Influence of a small background density}
\label{ssec:Background}

We now assume that $\on_B$ is nonzero but much smaller than $\on_m$
except for a narrow boundary layer where the sine of
(\ref{eq:psi-m-cg}) approaches zero. In that case \Eq{eq:rhoRatio1}
gives an upper bound to the ratio of densities at successive times,
that is very accurate in the interior of the system. Consequently, the
evaluation of \eq{eq:iirr2} carries over except that the boundary term
$\iirr_b$ picks up contributions due to the finite density $\on_B$ in
the cells $0$ and $N+1$. In the macroscopic limit this term becomes
\begin{equation} \label{eq:iirrb}
\iirr_b
 \to
   - \kappa (\ntot - \ntot^{(\infty)}) \left( 1 + \ln\frac{\on_B}{\rstar} \right)
 \approx
   - \kappa \ntot \left( 1 + \ln\frac{\on_B}{\rstar} \right) .
\end{equation}
Since the total entropy $S_{\rm tot}$ and the contribution
$\iirr_t$ to the entropy production do not
significantly change for a sufficiently small background density, the
full entropy production becomes in the macroscopic limit
\begin{equation}
   \iirr_{\rm relax}
  =
  - \Esc \ntot\;
     \ln\frac{\on_B}{\rstar} - \Esc  S_{\rm tot}  .
\label{eq:swbckgrd}
\end{equation}
After substituting $S_{\rm tot}$ from (\ref{eq:Stot}), we obtain the
specific irreversible entropy production
for ${\cal N} \gg {\cal N}^{(\infty)}$, 
\begin{equation}
  \specent^{\rm (irr)}
=
\left\{ \begin{array}{lrl}
\Esc
   \ln\left( \Pe^{-1} \frac{\ntot}{\ntot^{(\infty)}} \right)
   & \qquad {\rm for } & \Pe\gg 1,
\\[2mm]
  \Esc \ln\left( \frac{\ntot}{\ntot^{(\infty)}} \right)
   & {\rm for } & \Pe\ll 1.
\end{array}\right.
\label{eq:SigTotBackground}
\end{equation}
The result clearly shows that the logarithmic divergences in the
entropy production of the previous case are due to the vanishing
of a physically indispensable background density $\on_B$.

\section{Discussion}
\label{sec:Discussion}

The result \eq{eq:SigTotBackground} involves only well-behaved
macroscopic quantities, and the logarithm of the ratio of the number
$\ntot$ of particles in the system over the number
$\ntot^{(\infty)}=\rho_B L$ of particles approached in the steady
state. At intermediate times, where $\ntot$ is still much larger than
$\ntot^{(\infty)}$, the ratio $\ntot/\ntot^{(\infty)}$ decreases to a
good approximation exponentially like $\exp(-\Esc t)$ such that the
rate of irreversible entropy production starts to decrease linearly
like $-\Esc^2 t$.  During this time regime the boundary contribution
is by a factor of $\log(\ntot/\ntot^{(\infty)})$ larger than the bulk
contributions accounted for by \Eq{eq:sDotIrr}. Hence, even in the
more realistic setting accounting for a finite background density,
\Eq{eq:sDotIrr} only describes a sub-dominant contribution to the
entropy-production rate.
The reason for its failure is that the contributions arising from the
spatial distribution of the particles and the induced inhomogeneous
particle currents are not adequately taken into account by an entropy
based solely on the conditionally-invariant measure.

In spite of the strong contributions due to the boundary terms,
however, the entropy production remains proportional to the escape
rate $\Esc$ even in more general situations. Besides for the short
times, where this condition follows from \Eq{eq:SigTotBackground},
this can be easily illustrated in the long-time regime, for which
$(\ntot-\ntot^{(\infty)}) \psi \equiv \Delta\ntot \psi \ll \rho_B$.
Consequently, the entropy-production rate is obtained as
\begin{eqnarray}
   \Sigma^{\rm (irr)} 
&=&
   \int {\rmd x} \; \frac{\on(x)}{D} \;
   \left( v - D \frac{\Delta\ntot\; \partial_x\psi}{\on(x)} \right)^2
\nonumber \\
&\approx&
     \frac{v^2}{D} \; \int \rmd x \; \on(x)
   - \frac{\Delta\ntot^2}{\on_B}\, D \; 
     \int {\rmd x} \; \psi(x) \; \partial^2_x\psi(x) 
\nonumber \\
&\approx&
   \ntot \frac{v^2}{D} 
 + \Esc \; \frac{(\Delta\ntot)^2}{\ntot^{(\infty)}}  \; g(\Pe)
\label{eq:discussion}
\end{eqnarray}
where $g(\Pe) = \int\rmd x \, \psi^2(x)$ is a function of $\Pe$ only. 
The corresponding irreversible entropy production due to relaxation is
$\Sigma^{\rm (irr)}_{\rm relax} = \Sigma^{\rm (irr)} -  \ntot \, v^2/D$.
In the second line in \Eq{eq:discussion} the term proportional to
$\partial_x \psi$ does not appear since its integral vanishes.
Moreover, an integration by parts was used to obtain a second spatial
derivative of the density which, according to the advection-diffusion
equation, is proportional to its time-derivative, \ie it amounts to
$-\Esc \psi$ for the slowest decaying mode $\psi$ (again it is used
here that terms proportional to $\partial_x \psi$ and $\psi\,
\partial_x \psi$ vanish under the integral). 
The function $g(\Pe)$ can easily be evaluated for the multibaker
map, but in general it depends on the shape of the system.
Thus, the specific irreversible
entropy production taken with respect to $\Delta {\cal N}$ is
\begin{equation}
P^{\rm (irr)} =
\frac{\Sigma^{\rm (irr)}_{\rm relax}}{\Delta {\cal N}}
=
\kappa \frac{\Delta {\cal N}}{{\cal N}^{(\infty)}} g(\Pe) .
\label{eq:disc2}
\end{equation}
Equation \eq{eq:disc2} implies that even in a general thermodynamic
setting the relaxational entropy-production is proportional to the
escape rate $\Esc$, which characterizes the approach towards the
stationary state. In contrast to the dynamical-system arguments
\cite{BTV96} based on the escape-rate formalism, the term involves in
general a non-trivial function $g(\Pe)$ of the P\`eclet number, and it
has an amplitude $\Delta\ntot/\ntot^{(\infty)}$ that is exponentially
decaying like $\exp(-\Esc t)$.

It will certainly be interesting to investigate more closely the
connection between the escape-rate formalism and the decay to
systems supporting non-trivial stationary states. Another first
step in this direction, which complements the present approach was
suggested by Gilbert et al \cite{GilDorGas00}, who recently
discussed the approach towards equilibrium in a system with
periodic boundary conditions.

\acknowledgments

We acknowledge useful discussions with Bob Dorfman, Henk van Beijeren,
and Pierre Gaspard.  The research was supported by the Hungarian
Research Foundation (OTKA T032423) and the Schloessmann Foundation of
the Max-Planck Society.


\end{document}